\documentclass[conference]{IEEEtran}
\IEEEoverridecommandlockouts

\usepackage{CJKutf8}
\usepackage{kotex} 

\usepackage{cite}
\usepackage{amsmath,amssymb,amsfonts}
\usepackage[ruled,vlined,linesnumbered]{algorithm2e}

\usepackage{graphicx}
\usepackage{textcomp}
\usepackage{xspace}
\usepackage{xcolor}
\usepackage{comment}
\usepackage{graphicx}
\usepackage{tabularx}
\usepackage{array}
\usepackage{colortbl}
\usepackage{listings}
\usepackage{multirow}
\usepackage{enumitem}
\usepackage{makecell}
\usepackage{enumitem}
\usepackage{caption}
\usepackage{subcaption}\usepackage{tabularx}
\usepackage{booktabs}
\usepackage{amssymb}
\usepackage{hhline}
\usepackage{ulem}
\usepackage{amsfonts}
\usepackage{amsmath}
\usepackage{mathtools}

\usepackage[hyphens]{url} 
\usepackage{hyperref}
\usepackage{cleveref}
\usepackage{url}
\usepackage{balance}
\usepackage{ragged2e}

\crefformat{section}{\S#2#1#3} 
\crefformat{subsection}{\S#2#1#3}
\crefformat{subsubsection}{\S#2#1#3}

\newcommand{\ourtool}[0]{{\sc ALPS}\xspace}

\newcommand{\blue}[1]{\textcolor{black}{#1}}

\def\BibTeX{{\rm B\kern-.05em{\sc i\kern-.025em b}\kern-.08em
    T\kern-.1667em\lower.7ex\hbox{E}\kern-.125emX}}
\begin{document}
\title{\ourtool{}: Automated Least-Privilege Enforcement \\ for Securing Serverless Functions}

\author{
\IEEEauthorblockN{Changhee Shin\thanks{* Changhee Shin and Bom Kim contributed equally to this work. (This work was conducted while Bom Kim was with Incheon National University.)
\newline\hspace*{1.5em} Seungsoo Lee (seungsoo@inu.ac.kr) is the corresponding author.
\newline\hspace*{1.5em} \copyright~2026 IEEE. Accepted for publication at IEEE INFOCOM 2026.}}
\IEEEauthorblockA{
\small
Incheon National University\\
Republic of Korea\\
changhee9149@inu.ac.kr
}\vspace*{-2em}
\and
\IEEEauthorblockN{Bom Kim}
\IEEEauthorblockA{
\small
KAIST\\
Republic of Korea\\
spring@kaist.ac.kr
}\vspace*{-2em}
\and
\IEEEauthorblockN{Seungsoo Lee}
\IEEEauthorblockA{
\small
Incheon National University\\
Republic of Korea\\
seungsoo@inu.ac.kr
}\vspace*{-2em}
}

\maketitle
\begin{abstract}
Serverless computing is increasingly adopted for AI-driven workloads due to its automatic scaling and pay-as-you-go model. However, its function-based architecture creates significant security risks, including excessive privilege allocation and poor permission management. In this paper, we present \ourtool{}, an automated framework for enforcing least privilege in serverless environments. Our system employs serverless-tailored static analysis to extract precise permission requirements from function code and a fine-tuned Large Language Model (LLM) to generate language- and vendor-specific security policies. It also performs real-time monitoring to block unauthorized access and adapt to policy or code changes, supporting heterogeneous cloud providers and programming languages. 
\blue{In an evaluation of 8,322 real-world functions across AWS, Google Cloud, and Azure, \ourtool{} achieved 94.8\% coverage for least-privilege extraction, improved security logic generation quality by 220\% (BLEU), 124\% (ChrF++) and 100\% (ROUGE-2), and added minimum performance overhead. These results demonstrate that \ourtool{} provides an effective, practical, and vendor-agnostic solution for securing serverless workloads.}

\end{abstract}

\begin{IEEEkeywords}
Serverless Computing, Function-as-a-Service, Least Privilege, Access Control
\end{IEEEkeywords}

\section{Introduction}
\label{s:intro}

Serverless computing, a cloud model that enables developers to run applications without managing server infrastructure, is rapidly evolving with the rise of generative AI. AI inference workloads fluctuate significantly based on time and user demand, making them difficult to handle with traditional fixed infrastructure. Serverless architectures, with their automatic scaling and usage-based billing, address this challenge and have led to the emergence of Serverless Inference, which applies serverless principles to AI/ML model deployment. 
\blue{Major providers, including Hugging Face~\cite{huggingfaceserverless}, AWS Lambda~\cite{awslambda}, Google Cloud Functions~\cite{gcpfunction}, Azure Functions~\cite{azurefunction}, and Alibaba Function Compute~\cite{alibabafunction}, now support this approach. }
As a result, the global serverless computing market is projected to grow at a compound annual rate of 20.9\%, from USD 121.6 billion in 2023 to USD 811 billion by 2033~\cite{marketus}.

However, those advancements introduce significant security challenges, particularly inadequate permission management and excessive authorization. Unlike traditional monolithic applications, serverless environments decompose business logic into numerous independent functions, each requiring specific IAM permissions. CheckPoint research indicates that 98\% of serverless functions are over-privileged, with 16\% posing critical security risks~\cite{checkpoint2024serverless}. Developers often rely on wildcards (*) or overly broad managed policies for convenience, further exacerbating the problem~\cite{sysdig2024}.

Meanwhile, permission management issues persist after deployment. 
According to Orca Security (2023), 82\% of organizations have IAM user credentials that haven't been used for at least 90 days, and 72\% have unused IAM roles~\cite{orca2023}.  
\blue{In the 2021 Sendtech data breach, an unrotated AWS access key created in 2015 allowed attackers to gain administrative access and steal personal data~\cite{sendtech2021}. }
The `set and forget' problem~\cite{barracuda2025iam}, where temporary privilege escalations are not revoked, is also widespread. 
In serverless environments, frequent function creation and deletion further increase the risk of indirect privilege abuse and make it impractical to maintain consistent security policies manually.

Several studies have attempted to address such challenges in serverless permission management. Static analysis approaches~\cite{gupta2025growlithe, polinsky2024grasp, shimizu2020test, gill2022least, alpernas2018secure, jegan2023guarding, datta2020valve, gu2025epscan},  identify security risks by examining function code or policy structures but cannot automatically derive the minimum permissions required, relying instead on manual policy creation. Log-based methods~\cite{d2024automatically, xu2023log2policy, satapathy2023disprotrack}, generate policies from historical network activity or access logs but require lengthy data collection and fail to adapt to new or unforeseen traffic patterns. Runtime monitoring techniques~\cite{shin2025bambda, shen2022gringotts, alpernas2021cloud, datta2022alastor} detect permission abuse or security breaches in real time but primarily function as post-mortem auditing tools, lacking immediate response capabilities. Overall, these approaches cannot automatically extract least privileges, identify excessive permissions, or detect mismatches between function code and assigned policies.

In this paper, we propose \ourtool{}, an integrated framework that enforces least privilege in serverless environments. \ourtool{} uses serverless-tailored static analysis to extract precise permission requirements from function code and a fine-tuned Large Language Model (LLM) to generate programming language- and vendor-specific security policies. It further monitors cloud service calls in real-time, blocking unauthorized access and adapting to policy or code changes. By supporting heterogeneous cloud vendors and runtime languages, \ourtool{} provides a unified and automated approach to secure serverless function execution.

In the evaluation, \ourtool{} demonstrates its effectiveness and practicality across diverse serverless environments. In a least-privilege extraction experiment involving 8,322 serverless functions across AWS, Google Cloud, and Azure, our system achieved a coverage of 94.8\%, confirming its applicability to real-world workloads. 
\blue{For security verification code generation, the fine-tuned LLM improved BLEU scores by 220\%, ChrF++ scores by over 124\% and ROUGE-2 scores by over 100\% compared to the baseline, highlighting its ability to produce accurate, serverless-specific security logic.} Additionally, the performance overhead remained low, with execution times increasing by an average of 3.9ms on AWS, 5.5ms on Google Cloud, and 7.2ms on Azure, while cost increases across all platforms were minimal. These results indicate that \ourtool{} is both an effective and efficient solution for least-privilege management in production-grade serverless environments.

This paper makes the following contributions:

\begin{itemize}
\item We present the design and implementation of \ourtool{}, a fully automated and vendor-agnostic framework for extracting and enforcing least privilege in serverless environments.

\item We present an intelligent and adaptive verification code integration approach leveraging a fine-tuned LLM, enabling the system to monitor all access attempts and prevent privilege abuse at runtime without manual intervention.

\item We present automated policy conversion that translates uniform security requirements into vendor-specific formats, ensuring consistent security across heterogeneous cloud environments.

\item We evaluate \ourtool{} across major cloud platforms and real-world serverless workloads, demonstrating high permission extraction accuracy, robust runtime enforcement, and minimal performance overhead.
\end{itemize}

\section{Background and Problem Statements}
\label{s:back}

\subsection{Background}

\begin{figure}[t]
    \centering
    \includegraphics[width=0.9\columnwidth]{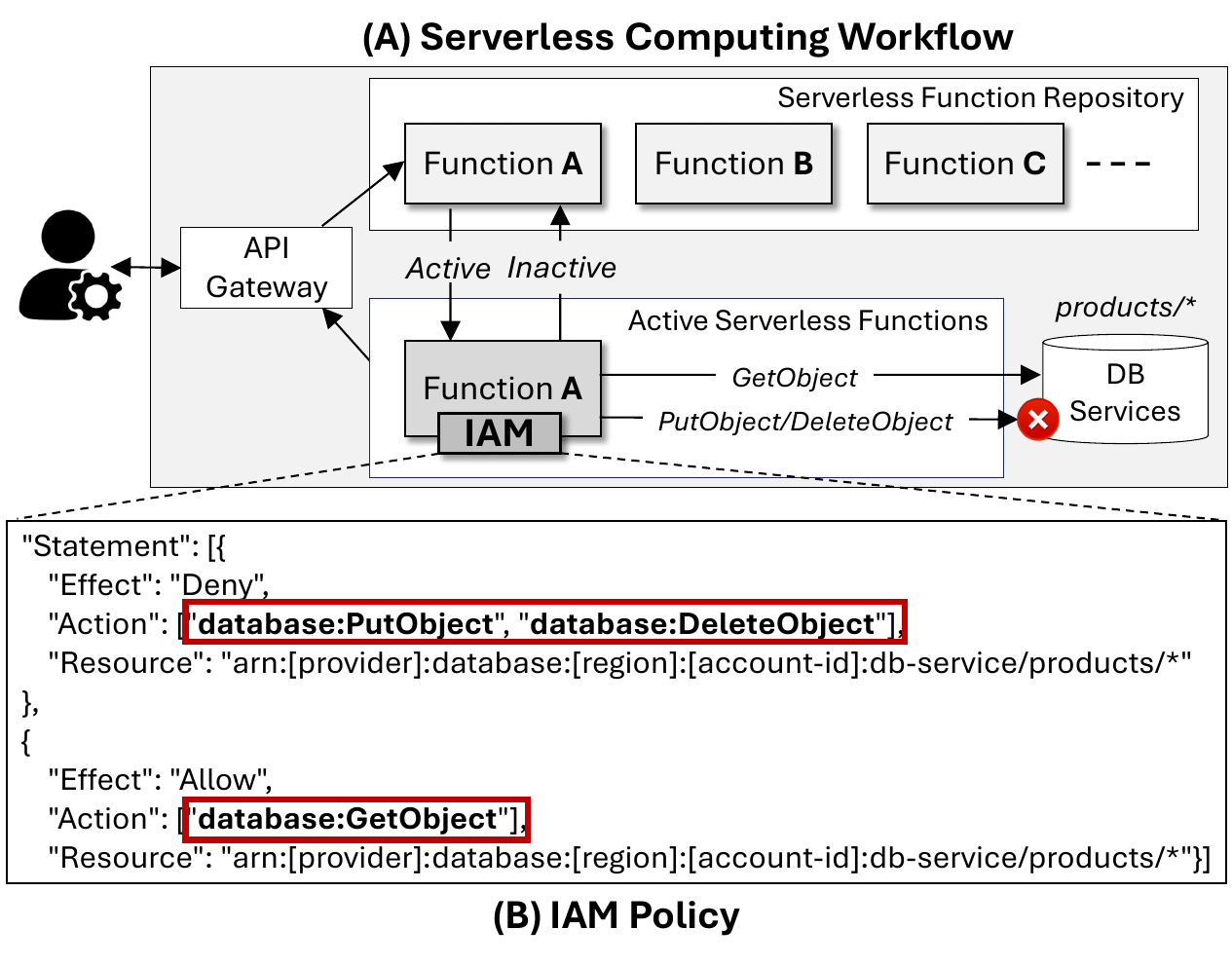}
    \caption{The example of serverless computing workflow and general IAM-based access control.}
    \label{fig:serverless-example}
\vspace{-1.3em}
\end{figure}

\textbf{Serverless Function Invocation.}
Serverless computing~\cite{li2022serverless, shafiei2022serverless, hassan2021survey} is an event-driven model where code executes at the function level, allowing developers to focus solely on business logic while cloud providers automate server provisioning, scaling, and maintenance through Function as a Service (FaaS). \blue{Popular platforms include AWS Lambda~\cite{awslambda}, Google Cloud Functions~\cite{gcpfunction}, Azure Functions~\cite{azurefunction}, and Alibaba Function Compute~\cite{alibabafunction}. As shown in Figure~\ref{fig:serverless-example}(A),} user requests are routed via an API Gateway to the appropriate functions, which independently access cloud resources and return results. \blue{These functions are stateless and short-lived, terminating after execution, which enables a pay-as-you-go pricing model~\cite{serverlesspricing} where costs apply only to actual function runtime.}

Functions are triggered via \textit{API}, \textit{Direct}, or \textit{Event-driven} methods, each with specific security requirements. First, API Invocation is triggered via HTTP requests routed through API Gateways or load balancers, commonly used in user authentication, data retrieval, and real-time processing. Next, Direct Invocation explicitly calls another function, facilitating workflow composition and microservice inter-communication. Finally, Event-driven Invocation responds automatically to state changes or external events (e.g., file uploads, database changes, message queues), suitable for batch processing, data pipelines, and real-time stream processing. Each method involves unique security implications, requiring fine-grained access controls to ensure secure interaction between functions and cloud resources.

\textbf{Identity and Access Management (IAM).}
Identity and Access Management (IAM)~\cite{singh2023iam} is an access control system responsible for managing permissions that enable serverless functions to interact with cloud resources. Due to their stateless nature, serverless functions cannot maintain permanent credentials. Consequently, each function invocation dynamically assigns temporary credentials based on predefined IAM policies, granting controlled access to resources. IAM enforces fine-grained access control by associating the entity requesting permissions (Identity), the defined permission rules (Action), and the targeted resource (Resource).

Figure~\ref{fig:serverless-example}(B) demonstrates IAM policy enforcement controlling resource access for a function within an AWS Lambda environment. When Function A attempts database access, IAM evaluates the assigned policies to determine permission grants or denials. In the depicted scenario, two IAM policies are applied: the first explicitly denies data insertion (\texttt{database:PutObject}) and deletion (\texttt{database:DeleteObject}), while the second explicitly allows data retrieval (\texttt{database:GetObject}). As a result, the function is permitted to read database contents upon invocation but is restricted from inserting or deleting data, illustrating real-time IAM policy enforcement.

\subsection{Problem Statements}
\label{challenges}

Enforcing the principle of least privilege in serverless environments presents significant challenges due to vendor-specific architectural differences, varying resource access requirements across individual functions, and dynamic changes in permission configurations during runtime. This section examines three critical challenges associated with enforcing effective least privilege in serverless function environments.

\begin{figure}[t!]
    \centering
    \includegraphics[width=\columnwidth]{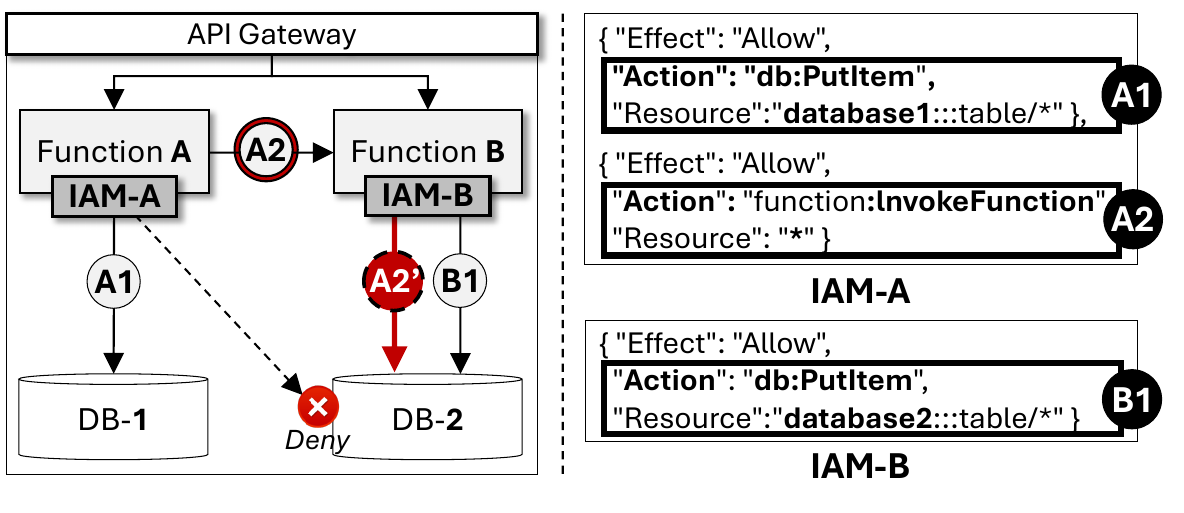}
    \caption{The motivating example of unauthorized permission abuse through function invocation bypass (A2 $\rightarrow$ A2') caused by the excessive permission (A2).}
    \label{fig:motivate}
\vspace{-1.4em}
\end{figure}

\textit{C1: Risk of Function Privilege Abuse Due to Manual Policy Management.}
Developers often assign uniform or global policies to multiple functions without fully analyzing their specific resource requirements, and even when tailored policies are defined, they frequently rely on overly simplified structures or wildcard permissions that grant unrestricted access across functions and resources~\cite{polinsky2024grasp}. Also, managed policies provided by cloud vendors further aggravate this issue by including unnecessary permissions, as demonstrated by the vulnerability in the \texttt{AmazonGuardDutyFullAccess} policy, which enabled organization-wide compromise~\cite{cymulate2025aws}. Although fine-grained permissions can be manually configured, the expertise and effort required make this approach impractical in large-scale serverless applications, thereby increasing the risk of privilege abuse.

Figure~\ref{fig:motivate} illustrates how overly permissive policy allocation can expose serverless applications to indirect privilege escalation through function invocation chains or event-driven interactions. Function A has permissions limited strictly to DB-1, while Function B can access DB-2. Despite individual IAM policies effectively restricting direct resource access (A1, B1), excessive invocation permissions enable Function A (and thus the invoking user) to indirectly access unauthorized resources, such as DB-2 (A2'), by invoking Function B. Although each IAM policy remains individually compliant, excessive invocation permissions enable users to circumvent intended access constraints. Such issues occur when developers fail to implement fine-grained permission policies that restrict inter-function invocations to explicitly defined and necessary scenarios.

\textit{C2: Lack of Runtime Enforcement of the Principle of Least Privilege.}
In serverless environments, even if the principle of least privilege is correctly applied during initial deployment, privileges are often expanded or modified during operation without proper revocation, leading to the Set-and-Forget problem~\cite{barracuda2025iam}. Existing permission verification mechanisms primarily rely on post-deployment analysis, which cannot prevent privilege abuse in real time~\cite{d2024automatically, shen2022gringotts}. While cloud vendor monitoring tools can detect abnormal or unauthorized privilege usage through log analysis, they lack function-level blocking capabilities~\cite{iamAccessAnalyzer, policyAnalyzer, msentra}. Similarly, policy-based approaches verify conditions at a logical level but do not enforce real-time, application-level checks to ensure resource access remains within permitted scopes. This absence of runtime verification allows privilege abuse to go undetected or unmitigated, significantly increasing the risk of security breaches.

\begin{figure*}[t!]
    \centering
    \includegraphics[width=0.91\textwidth]{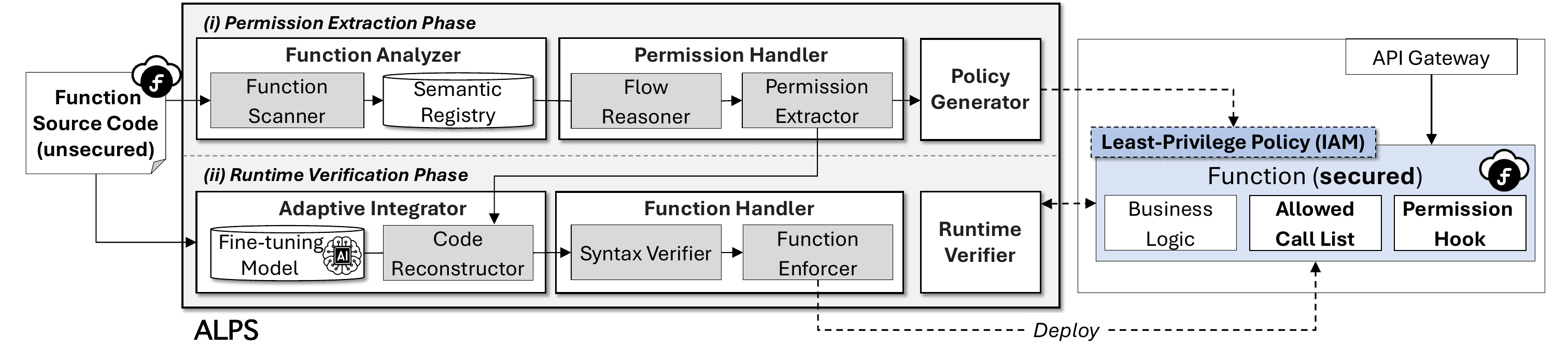}
    \caption{Overall architecture and its workflow of \ourtool{} with six key components: (\textit{i}) function analyzer, (\textit{ii}) permission handler, (\textit{iii}) policy generator, (\textit{iv}) adaptive integrator, (\textit{v}) function handler and (\textit{v}) runtime verifier. Additionally, our system includes two operational phases: permission extraction and runtime verification.}
    \label{fig:archi}
\vspace{-1.3em}
\end{figure*} 

\textit{C3: Lack of Consistency Across Cloud Providers in Serverless Environments.}
Each cloud provider uses a distinct IAM policy structure and permission management model, requiring developers to learn different syntax and configuration approaches for each platform~\cite{marin2022serverless, csa}. Serverless functions are also implemented in multiple programming languages, such as JavaScript, Python, and Go, further complicating the consistent enforcement of security policies across diverse language–platform combinations. Existing automated policy generation tools~\cite{orca} are typically tightly coupled to specific vendors or languages, making them unsuitable for deployments across heterogeneous cloud providers. Consequently, maintaining uniform security policies across heterogeneous environments demands separate tools and processes for each platform, increasing management complexity and hindering effective security governance for serverless applications.

\section{\ourtool{} Design}
\label{s:design}

This section outlines the design considerations that motivated the development of \ourtool{} and presents its system architecture. Succinctly, our system enhances the security of serverless functions by automatically extracting least-privilege permissions through static analysis and verifying their correct operation under these permissions at runtime.

\subsection{Design Considerations}
\textbf{1) Automatic Extraction of Fine-Grained Permissions for Enforcing Least Privilege.} The system should accurately identify and extract the minimum permissions required for each serverless function based on its specific tasks. This requires automatically detecting cloud service calls and the corresponding resources accessed through function code analysis and deriving precise permission requirements using data flow inference. Furthermore, even in complex workflows involving direct invocations or event-driven function chaining, it should differentiate fine-grained permissions to ensure that each function is granted only the privileges necessary for its role, without manual developer intervention.

\textbf{2) Real-Time Permission Monitoring and Abuse Prevention.} The system should continuously ensure that functions operate within the minimum required privilege level after deployment. It should validate the legitimacy of all SDK service calls before execution using a dynamically generated whitelist derived from extracted minimum privileges and environment variables, immediately blocking any unauthorized access attempts. Furthermore, when function code or policies are modified, it should re-evaluate the function to update the whitelist, detect excessive privileges, and enforce appropriate execution controls.

\textbf{3) Intelligent Integration Across Cloud Vendors and Programming Languages.} The system should provide consistent authorization management across heterogeneous environments with varying policy grammars and code structures. This requires context-aware, intelligent code generation that goes beyond predefined template mappings by analyzing coding styles, variable naming patterns, and exception handling methods in real time to insert authorization verification logic without disrupting the existing program structure. Furthermore, the system should translate uniform security requirements into vendor-specific IAM policy grammars, enabling seamless management of heterogeneous cloud environments without additional configuration overhead for developers.

\begin{figure*}[t!]
    \centering
    \includegraphics[width=0.91\textwidth]{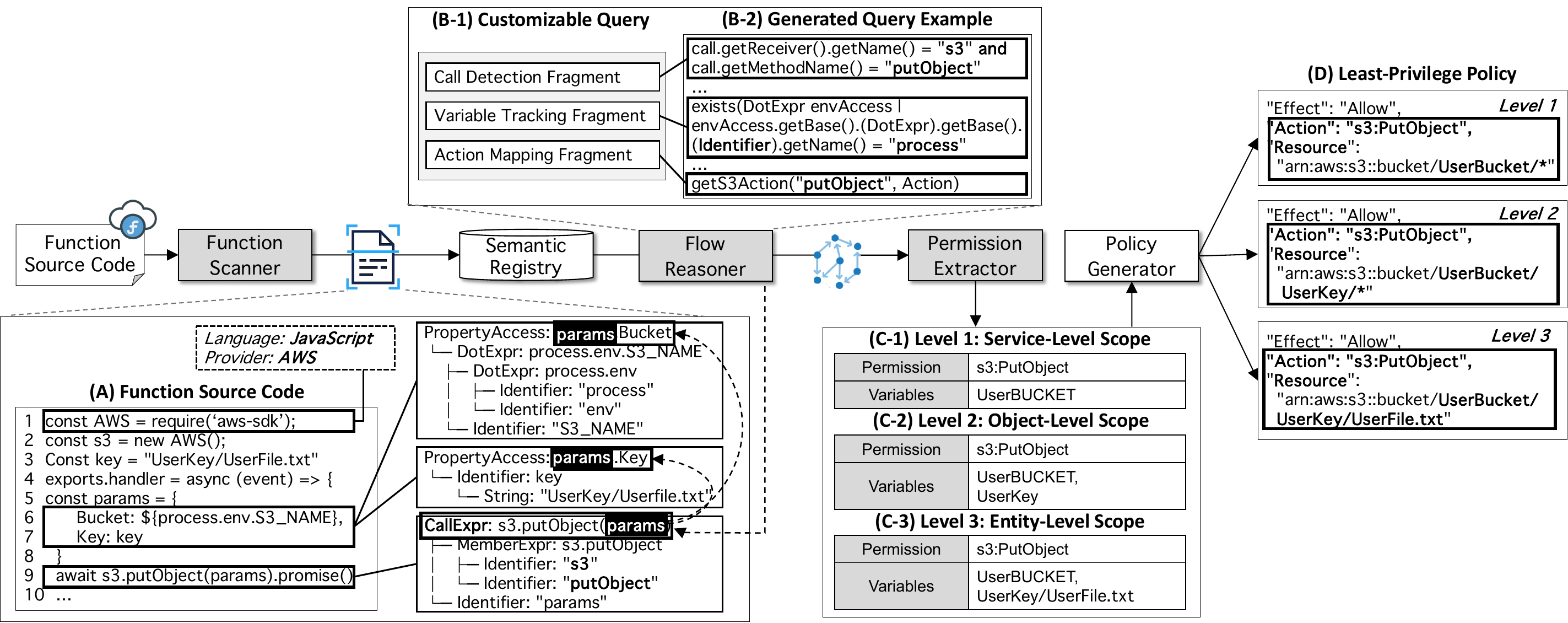}
    \caption{The example of automatic extraction procedure of least-privileged policies and environment variables.}
    \label{fig:policy-generation}
\vspace{-1.4em}
\end{figure*}

\subsection{System Architecture and Workflow}
To satisfy the design considerations outlined above, \ourtool{} (depicted in Figure~\ref{fig:archi}) comprises six core components: function analyzer, permission handler, policy generator, adaptive integrator, function handler, and runtime verifier. Our system accepts only the source code of serverless functions as input, without requiring any additional configuration files or metadata. The input functions represent typical serverless functions containing only standard business logic, without any security-specific modifications. Based on these inputs, the system operates in two stages: permission extraction phase and runtime verification phase, as described below.

First, the permission extraction phase automatically identifies the minimum permissions required for each serverless function and generates the corresponding policies. The \textit{function analyzer} processes the source code, detects the programming language, and produces a structured code representation through static analysis, which is stored in the semantic registry as a knowledge base for subsequent analysis. Using this information, the \textit{permission handler} detects cloud service calls and resource access paths, constructing a complete data flow that incorporates conditional branching, dynamic resource access, and environment variable references. Based on this flow, the permission extractor derives the minimal permission set (i.e., actions and resources) needed by the function, along with the resolved environment variable values, which are later used to create the allowed call list. Finally, the \textit{policy generator} translates the extracted permissions into the appropriate IAM policy format for each cloud vendor.

Next, the runtime verification phase ensures that deployed functions operate strictly within their minimum required permissions while dynamically adapting to IAM policy changes after deployment. The \textit{adaptive integrator} employs a fine-tuned Large Language Model (LLM), trained on a curated code integration dataset, to insert allowed call lists and permission monitoring hooks customized for each programming language and cloud provider. Guided by the code reconstructor, the LLM integrates extracted permissions and environment variables into the original function code, producing a secure, instrumented version. The \textit{function handler} then validates the reconstructed code for syntactic correctness and functional equivalence before deploying it.

During execution, the \textit{runtime verifier} inspects all SDK service calls in real-time. Each call is validated against a dynamically generated whitelist derived from environment variables, and any unauthorized access attempt is immediately blocked before execution. If function code changes, the updated code is reanalyzed in the permission extraction phase, and the whitelist is regenerated. \blue{Similarly, if an IAM policy change is detected, the system compares the updated policy against the whitelist and enforces execution within the scope of the whitelist, even if the policy grants excessive privileges.}

\section{\ourtool{} System Details}
\label{s:details}

\subsection{Automated Least-Privilege Extraction via Serverless-Tailored Static Analysis}
\label{ss:4-A}
In serverless environments, function-level source code is deployed directly to the cloud, granting developers full visibility and access. Each function has well-defined inputs, outputs, and logic, constraining its tasks and resource access. These properties allow static analysis to accurately identify execution flows and resource access patterns. Figure~\ref{fig:policy-generation} illustrates how \ourtool{} extracts least-privilege permissions.

\textbf{Serverless-Tailored Static Analysis.} 
Existing general-purpose static analysis tools, such as CodeQL~\cite{codeql} offer robust Abstract Syntax Tree (AST) generation and basic dataflow tracing across multiple languages. However, the complexity of cloud SDK services and the variability in call structures across vendors hinder their ability to accurately determine a function's precise permission requirements. To address this limitation, we extend the infrastructure of the CodeQL with a serverless-specific query framework tailored to accurately identify and trace cloud service calls.

\textit{(i) Language and Vendor Identification:} The analysis begins with the function scanner automatically identifying the programming language and cloud vendor from the function source code ((A) in Figure~\ref{fig:policy-generation}). Language detection leverages file extensions, import statements, and syntax patterns, while vendor identification relies on analyzing SDK imports and client instantiation patterns. These language–vendor mappings form the basis for selecting the appropriate queries and rule sets for subsequent analysis.

\textit{(ii) AST-Based Structuring:} Based on the identified language, CodeQL is employed to transform the function code ((A) in Figure~\ref{fig:policy-generation}) into an AST representation, enabling precise decomposition and structuring of cloud service call syntax. For example, instead of treating a cloud SDK call (\texttt{s3.putObject(params)}) as a single function call, it is decomposed into AST nodes that separately capture the service client (\texttt{s3}), method name (\texttt{putObject}), and parameters (\texttt{params}). Similarly, environment variable references (\texttt{process.env.S3\_NAME}) and resource identifier accesses (\texttt{params.Bucket}, \texttt{params.Key}) are represented as individual nodes, allowing the tracking of dynamic resource reference patterns. The resulting semantic registry structures each component of a cloud service call into AST nodes and enables data flow tracking by automatically linking these nodes during query execution. This allows for more precise identification of authorization requirements in subsequent steps.

\textit{(iii) Automatic Query Building and Execution:} To accurately identify and track cloud service calls, we develop a customizable query system ((B-1) in Figure~\ref{fig:policy-generation}) tailored to each language–vendor combination. This system comprises three main query fragments. The Call Detection Fragment detects service invocations ranging from direct cloud SDK calls to indirect calls made through wrapper functions or utility modules. The Variable Tracking Fragment traces the data flow of variables relevant to authorization decisions, including environment variable references (\texttt{process.env.BUCKET\_NAME}), configuration file loading, and dynamically generated resource identifiers. Finally, the Action Mapping Fragment defines mapping rules that translate identified SDK method calls into specific IAM actions for the corresponding cloud vendor.

Based on the established query structure, the flow reasoner automatically generates the analysis queries optimized for the identified language–vendor combination (B-2). For example, for the AWS–JavaScript combination, it generates a query that detects the \texttt{putObject(UserBUCKET, UserKEY, putParams)} call and traces the data flow of the \texttt{UserBUCKET} and \texttt{UserKEY} variables. When executed on the semantic registry, the flow reasoner derives cloud service call points and their associated data flows. Through this process, it establishes the connection between the \texttt{putObject} call and the \texttt{UserBUCKET} and \texttt{UserKEY} variables, identifies dynamic resource reference patterns (e.g., environment variables), and determines all resources and actions accessible to the function during execution. This analysis forms the basis for the subsequent least-privilege extraction step.

\textbf{Least Privilege Extraction.}
The permission extractor determines the minimum permissions required by a function based on the data flow derived by the flow reasoner. For each identified cloud service call point, it applies the Action Mapping Fragment rules to map method calls (\texttt{putObject}) to their corresponding IAM actions (\texttt{s3:PutObject}) and associates variables (\texttt{UserBUCKET}, \texttt{UserKEY}) with the corresponding bucket and object resources. For calls within conditional branches or loops, it analyzes all possible execution paths to ensure comprehensive permission coverage.

\blue{Based on the extracted permissions, the system provides three granular levels of access: (i) \textit{Service-Level Scope} (C-1 in Figure~\ref{fig:policy-generation}), which grants only service-specific action permissions; (ii) \textit{Object-Level Scope} (C-2), which additionally specifies key resource identifiers; and (iii) \textit{Entity-Level Scope} (C-3), which produces the most fine-grained permissions, encompassing all relevant variables and execution paths.} This approach minimizes the use of wildcards while allowing users to select the appropriate permission level based on operational and management requirements. Runtime enforcement, however, adheres to the most granular level (C-3) following the principle of least privilege.

\blue{Finally, the policy generator automatically translates the extracted permissions into vendor-specific IAM policy formats, including AWS, Google Cloud, and Azure, producing ready-to-deploy minimum privilege policies (D in Figure~\ref{fig:policy-generation}).}

\subsection{Real-Time Monitoring and Verification of Permission Abuse}
\label{ss:4-B}

\ourtool{} implements a mechanism for intercepting and verifying all cloud service calls made by a deployed serverless function in real-time. This verification is achieved through permission hooks embedded in the function code, enabling continuous monitoring of all cloud service calls during execution and automatically accommodating configuration changes that occur after deployment. These hooks operate based on an allowed call list of permission–resource pairs, ensuring that it remains non-intrusive to the business logic of the function. The process for generating the allowlist and integrating permission hooks into the function code is described in Section~\ref{ss:4-C}.

\LinesNotNumbered

\begin{algorithm}[t]
\small
\DontPrintSemicolon
\caption{\blue{Hierarchical Verification for Service Calls.}}
\label{verification_algorithm}
\KwIn{Service Calls Event $E$}
\KwOut{0(Allow) or 1(Deny)}

action $\gets$ get\_current\_action(E)\;

service $\gets$ get\_current\_service(E)\;

\If{requires\_resource\_extraction(E)}{
    resource $\gets$ extract\_resource\_name(E, service)\;
}
\Else{
    resource $\gets$ default\_resource\_identifier(E)\;
}

allowlist $\gets$ get\_allowed\_calls()\;

is\_allowed $\gets$ FALSE\;

\If{service $\in$ allowlist}{
    allowed\_resource $\gets$ allowlist[service]\;
    
    \If{resource $\in$ allow\_resource}{
        allowed\_action $\gets$ allowed\_resource[resource]\;
        
        \If{action $\in$ allowed\_action}{
            is\_allowed $\gets$ TRUE\;
        }
    }
}

\If{is\_allowed == FALSE}{
    \Return 1\;
}

\Return 0\;
\end{algorithm}

The verification process begins by intercepting every service request through hooks in the core service mechanisms of the cloud SDK. For example, in the Azure SDK, the request of the Cosmos DB client method is intercepted to capture database calls before execution. From each call, the service, operation, and parameters are extracted, and the target resource is dynamically resolved. For Cosmos DB, database and container names are analyzed, and if resource identifiers are environment variables, their values are retrieved at runtime. The identified service–resource pair and operation are then checked against the allowlist of the function. Unauthorized requests raise an exception and are blocked immediately. These steps are summarized in Algorithm~\ref{verification_algorithm}.

\textbf{Automated Handling of Post-Deployment Configuration Changes.} 
In serverless environments, various configuration changes can occur after deployment. \ourtool{} automatically adapts to these changes, ensuring continuous compliance with the principle of least privilege. When the source code of a deployed function is updated, the updated function code is reintroduced into the static analysis stage, where new permission sets and environment variables are derived, and the allowlist is regenerated.

When an IAM policy is modified, our system determines whether excessive permissions have been introduced. Even if additional permissions are granted, our system enforces execution strictly within the previously extracted least-privilege scope. 
\blue{By comparing the permissions defined in the new  policy with the current allowlist, \ourtool{} identifies over-permissive resources or actions and restricts function execution.}

\subsection{LLM-Based Adaptive Code Integration}
\label{ss:4-C}

\begin{table}[t]
\centering
\renewcommand{\arraystretch}{1.3}
\caption{The summary of the datasets for LLM fine‑tuning
         (Fn = Function,\; FT = Fine‑tuning Set).
         Origin function templates are sourced from the \textit{Wonderless} dataset~\cite{eskandani2021wonderless}.}
\label{tab:dataset}
\resizebox{\columnwidth}{!}{%
\begin{tabular}{l|ccc|ccc|cc}
\Xhline{3\arrayrulewidth}
\rowcolor{gray!15}
\cellcolor{white} &
\multicolumn{3}{c|}{\textbf{AWS}} &
\multicolumn{3}{c|}{\textbf{Google Cloud}} &
\multicolumn{2}{c}{\textbf{Azure}} \\  
 & JS & Py & Go & JS & Py & Go & JS & Py \\ \hline \hline
Function           & 5,475 & 889 & 39   & 9   & 21 & 16  & 31  & 22 \\ \hline 
Permission        & \multicolumn{3}{c|}{680} & \multicolumn{3}{c|}{676} & \multicolumn{2}{c}{660} \\ \hline
Curated Fn          & 30    & 14  & 24   & 9   & 10 & 9   & 15  & 10 \\
Refined FT          & 20,000 & 8,940 & 4,130 & 6,100 & 6,800 & 5,500 & 9,300 & 6,700 \\
\Xhline{3\arrayrulewidth}
\end{tabular}}
\vspace{-2em}
\end{table}

To ensure consistent authorization verification across diverse cloud vendors and programming languages in serverless environments, \ourtool{} introduces a fine-tuned, LLM-based adaptive code integration mechanism. Existing general-purpose code generation models often fail to capture the specialized authorization verification requirements of serverless functions or account for differences in SDK architectures across cloud providers. In addition, template-based approaches are limited in their ability to seamlessly integrate authorization logic into function code. To address these gaps, we fine-tuned an LLM on a domain-specific dataset for serverless authorization management, enabling it to learn the unique characteristics of each environment and perform context-aware code integration.

\textbf{Dataset Construction.} 
The fine-tuning dataset was built using serverless functions from the \textit{Wonderless} dataset~\cite{eskandani2021wonderless} and permission policy examples collected from GitHub and official documentation (Table~\ref{tab:dataset}). To ensure high-quality training data, we applied rigorous filtering criteria, including the removal of inactive projects (those with less than one year of activity), official community examples, duplicate code, and empty functions. From approximately 6,500 functions gathered across three cloud vendors, we excluded Azure functions in Golang due to the lack of suitable examples. From the remaining combinations, about 120 functions were selected for their representativeness and diversity, emphasizing common resource access patterns and function structures. \blue{Additionally, around 2,000 permission policy examples were evenly collected from AWS, Google Cloud, and Azure. The selected functions were curated and augmented using semantic-preserving transformations~\cite{zhang2025secon} to ensure clear function-permission mappings while generating diverse training scenarios.} This process yielded a high-quality dataset tailored to serverless permission management. The dataset was split into training and test sets (including validation) at an 8:2 ratio for reliable model evaluation.

\begin{figure}[t!]
    \centering
    \includegraphics[width=\columnwidth]{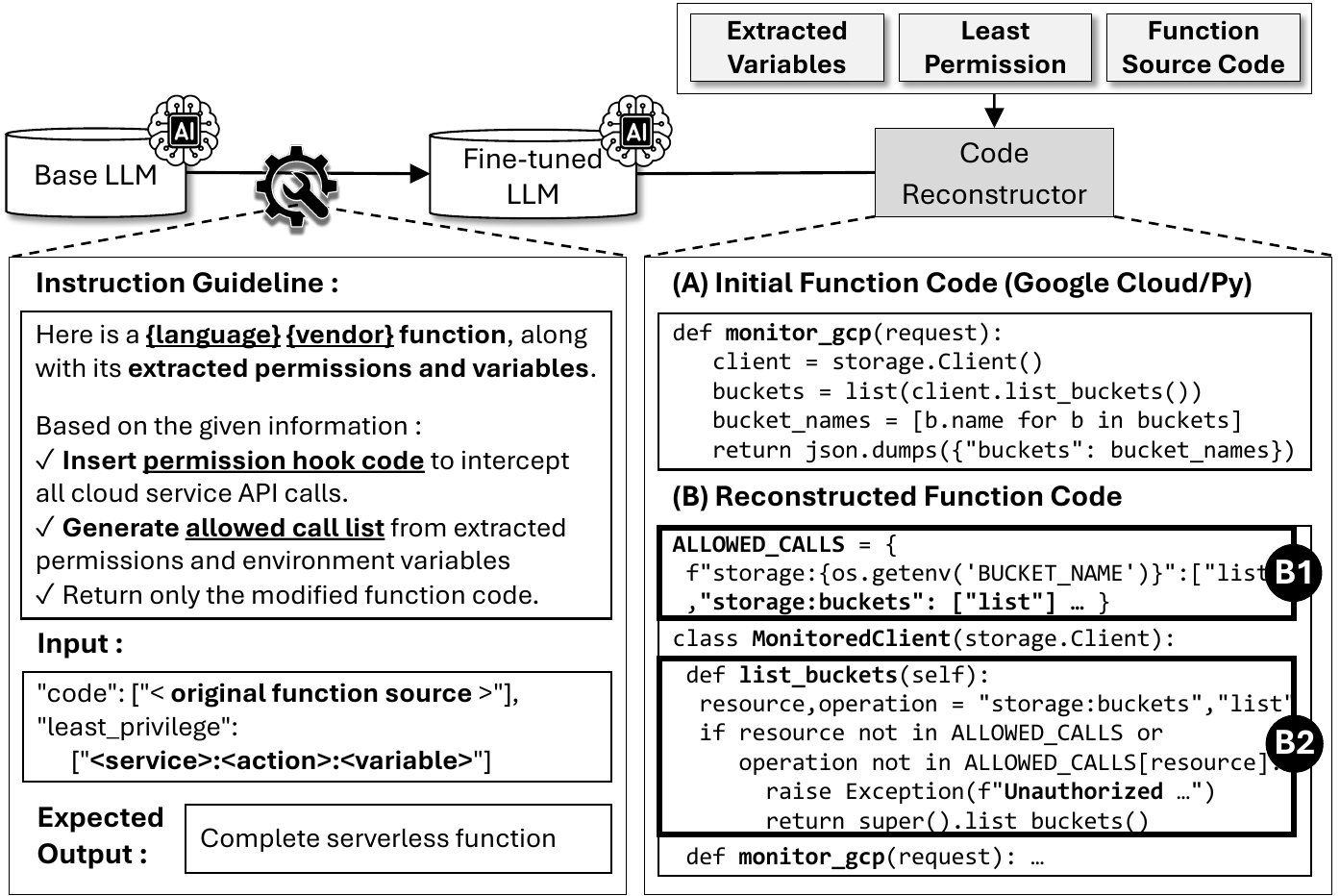}
    \caption{Instruction guideline for LLM fine-tuning and the example for automatic function restructuring by the code reconstructor.}
    \label{fig:code-integration}
\vspace{-1.2em}
\end{figure}

\textbf{LLM-Based Permission Verification Code Generation.}
As shown in Figure~\ref{fig:code-integration}, the Base LLM is adapted into a Fine-Tuned LLM via transfer learning that incorporates structured context information, enabling precise serverless permission verification code generation. The Code Reconstructor uses three inputs: (i) variables from static analysis, (ii) the least-privilege permission set, and (iii) the original function code. The model is trained to generate vendor- and language-specific functions, insert permission hooks, and build environment-aware allowlists, producing fully instrumented functions.

The trained model automatically integrates these components. In Figure~\ref{fig:code-integration}, the original function (A) only instantiated a Google Cloud Storage client, whereas the reconfigured function (B) includes permission verification logic. First, the model dynamically generates allowlists by combining environment variables with the derived permission set (B1). Second, it injects language-specific hooks: Python uses wrapper classes, JavaScript redefines SDK methods at runtime, and Go inserts explicit verification calls (B2). This approach enables automated, environment-optimized authorization enforcement with minimal developer intervention.

\section{Evaluation}
\label{s:eval}

\subsection{Implementation}

\blue{We implemented a prototype of \ourtool{} using CodeQL~\cite{codeql} and Python, comprising approximately 50,000 lines of code (primarily CodeQL queries, with $\sim$200 lines of core logic in Python), to evaluate its feasibility and effectiveness.}
The AST-based static analysis engine extracts resources, actions, and conditions from cloud service calls and infers the required permissions, which are then normalized into a unified representation compatible with AWS IAM~\cite{awsiam}, Google Cloud IAM~\cite{gcpiam}, and Azure RBAC~\cite{azurerbac}. Furthermore, we fine-tuned the DeepSeek 6.7B model~\cite{deepseek-ai} to automatically generate runtime security verification logic from function code and least-privilege specifications, incorporating platform- and language-specific characteristics.

\subsection{Evaluation Environments}
The experiments were performed on a high-performance server equipped with an NVIDIA Hopper H100 80GB PCIe GPU, an AMD EPYC 9224 CPU, 256GB RAM, and a 2TB SSD. We tested AWS Lambda~\cite{awslambda}, Google Cloud Functions~\cite{gcpfunction}, and Azure Functions~\cite{azurefunction} with Node.js, Python, and Go runtimes. The functions were collected from Wonderless~\cite{eskandani2021wonderless}, a benchmark dataset containing real-world serverless functions that span heterogeneous vendors and runtimes, and exhibit key serverless characteristics such as event chaining, database access, and external service calls. This setup enabled a comprehensive evaluation of the capability of our system to extract least privilege, generate security logic, and verify and enforce policies under realistic conditions.

\begin{figure}[t!]
    \centering
    \includegraphics[width=0.94\columnwidth]{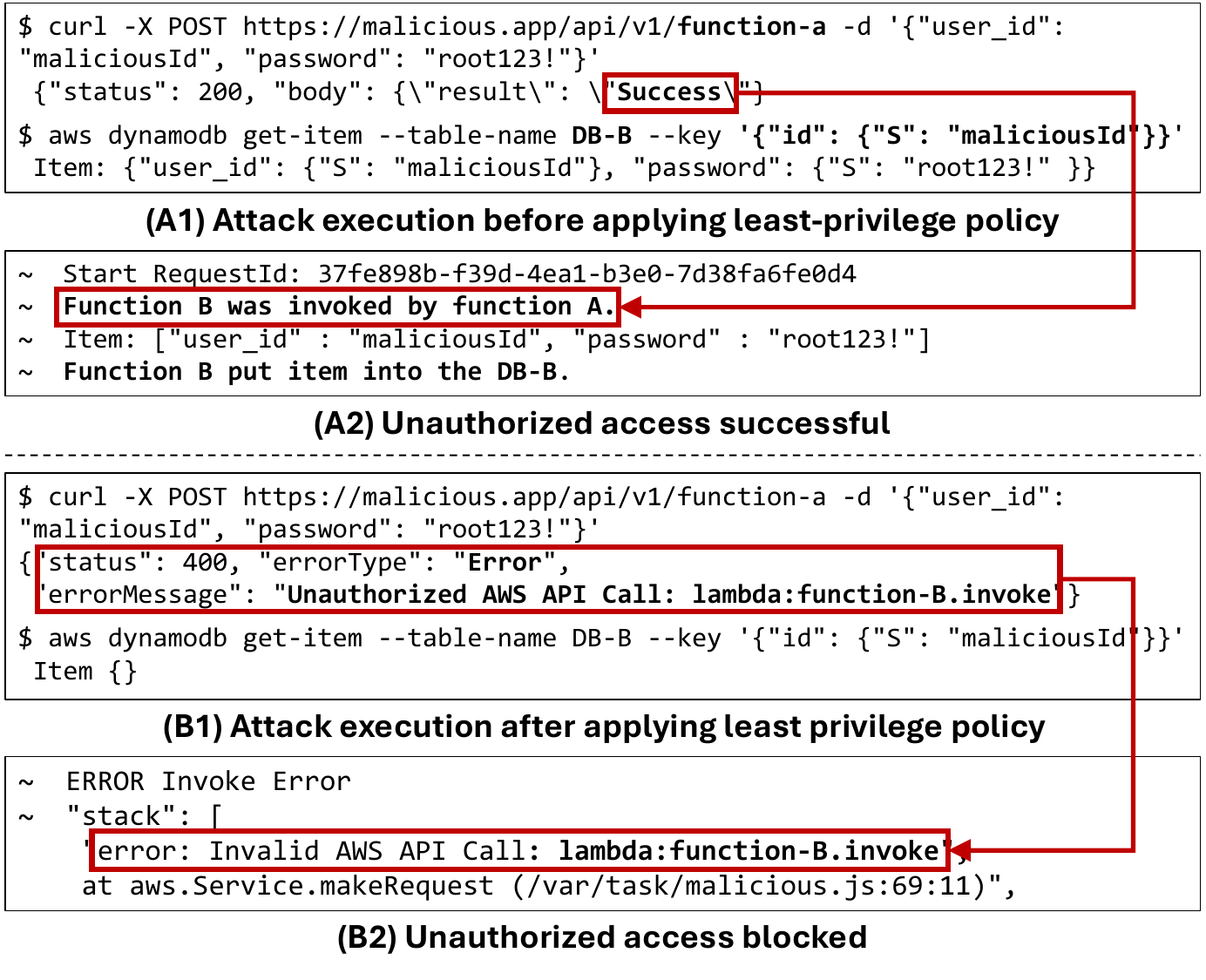}
    \caption{The comparison of the permission abuse results with and without least privilege policies generated by \ourtool{}.}
    \label{fig:leastIAM}
\vspace{-1.2em}
\end{figure}

\subsection{Functional Correctness}
This evaluation examines the effectiveness of \ourtool{} in enforcing least-privilege policies to prevent permission abuse. Figure~\ref{fig:leastIAM} compares the abuse outcomes before and after applying least-privilege policies. The experiment simulates an AWS Lambda environment where Function A is erroneously granted excessive permissions, including the ability to invoke other functions using wildcards, despite requiring access only to DB-A, as illustrated in Figure~\ref{challenges}. 
\blue{We constructed a permission abuse scenario in which Function A exploits these unnecessary privileges to invoke Function B and insert malicious data into DB-B, thereby assessing the defensive capability of \ourtool{}.}

While Function A successfully inserted data into DB-B by invoking Function B due to excessive permissions, as shown in Figure~\ref{fig:leastIAM} (A1-2), this attempt was promptly blocked after enforcing the least-privilege policy. \ourtool{} analyzed the function’s code, identified that it required only DB-A access permissions, and generated a policy that removed the unnecessary Lambda invocation privileges. The log in Figure~\ref{fig:leastIAM} (B1-2) further confirms the invocation failure and shows that the attempt to insert data into DB-B was blocked. These results demonstrate that our system accurately identifies actual resource access patterns through serverless-specific static analysis and generates least-privilege policies that effectively eliminate unnecessary permissions.

\subsection{Performance}

\begin{table}[t]
\centering
\scriptsize
\renewcommand{\arraystretch}{1.3}
\caption{Statistics of detection results per cloud platform and language. Origin Function templates are sourced from the \textit{Wonderless} datasets~\cite{eskandani2021wonderless}.}
\label{tab:coverage}
\resizebox{\columnwidth}{!}{%
\begin{tabular}{l|ccc|ccc|cc|c}
\Xhline{2.8\arrayrulewidth}
\rowcolor{gray!15}
\cellcolor{white} &
\multicolumn{3}{c|}{\textbf{AWS}} & 
\multicolumn{3}{c|}{\textbf{Google Cloud}} & 
\multicolumn{2}{c|}{\textbf{Azure}} & 
\cellcolor{white}\multirow{2}{*}{\textbf{Total}} \\
 & JS & Py & Go & JS & Py & Go & JS & Py & \\ 
\hline \hline 
Function      & 2512 & 1208 & 807 & 784 & 802 & 753 & 852 & 604 & 8322 \\ 
\hline
Detected    & 2393 & 1140 & 779 & 759 & 777 & 729 & 827 & 587 & 7891 \\
Undetected  &  119 &   68 &  28 &  25 &  25 &  24 &  25 &  17 &  331 \\
\Xhline{2.8\arrayrulewidth}
\end{tabular}}
\vspace{-1.0em}
\end{table}

\textbf{Function Coverage.}
We measure the coverage of permission extraction by \ourtool{} across real serverless workloads to assess the success rate of least-privilege extraction for diverse functions. Specifically, it aims to determine whether our system can achieve complete extraction at the \texttt{entity-level} scope, the most granular of the three permission granularity levels proposed. As shown in Table \ref{tab:coverage}, unlike the fine-tuning dataset, the evaluation dataset was constructed by collecting all functions containing cloud service calls from the \textit{Wonderless} dataset~\cite{eskandani2021wonderless} and removing duplicates and empty templates. This process yielded approximately 8,300 unique functions spanning three major cloud platforms (AWS, Google Cloud, and Azure) and three programming languages (JavaScript, Python, and Go).

As a result, \ourtool{} successfully extracted least-privilege policies for 7,891 functions, achieving 94.8\% coverage. Google Cloud had the highest success rate at 96.8\%, followed by AWS at 95.2\% and Azure at 90.3\%. Across languages, JavaScript, Python, and Go achieved 95.9\%, 95.8\%, and 90.3\% respectively, demonstrating consistent performance. The 331 functions (5.2\%) that failed entity-level extraction were mainly due to two static analysis limitations: dynamically determined resource identifiers (e.g., from service responses or user inputs) and service calls that require wildcard permissions (e.g., S3 ListBucket, Lambda ListFunctions). In these cases, only service-level permissions could be generated. These results confirm the broad applicability, vendor compatibility, and effectiveness of our system in extracting least-privilege policies across diverse serverless workloads.

\begin{figure}[t!]
    \centering
    \includegraphics[width=0.98\columnwidth]{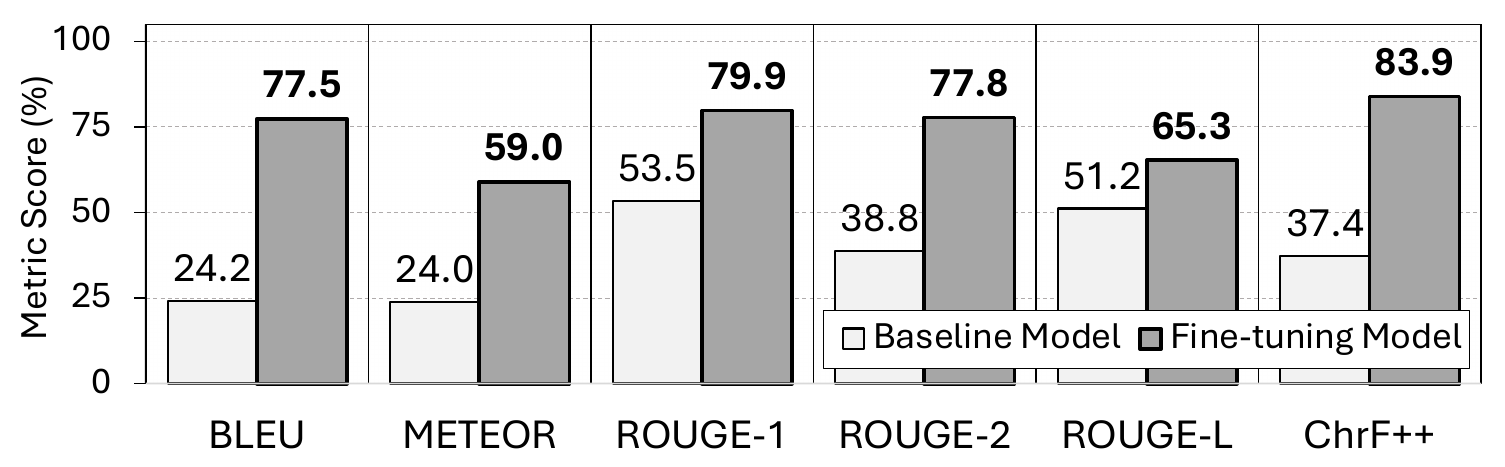}
    \caption{\blue{The summary of the fine-tuning LLMs performance with metrics: BLEU, METEOR, ROUGE-1, ROUGE-2, ROUGE-L, ChrF++.}}
    \label{fig:finetuning}
    \vspace{-1.3em}
\end{figure}

\textbf{Fine-tuning LLM Performance.}
To quantitatively evaluate the accuracy of verification logic generation in serverless environments, we compared the base model (DeepSeek Coder 6.7B Instruct~\cite{deepseek-ai}) with its fine-tuned version. For evaluation, we used a test set randomly sampled from 20\% of the dataset defined in Table~\ref{tab:dataset}. Each model generated function code with embedded verification logic based on the input function code and least-privilege specifications. 
\blue{Model performance was assessed using 
BLEU~\footnote{BLEU measures grammatical and syntactic similarity to the ground truth}~\cite{BLEU}, 
METEOR~\footnote{METEOR evaluates semantic alignment and word-level precision with synonym matching}~\cite{banerjee2005meteor}, 
ROUGE (ROUGE-1, ROUGE-2, ROUGE-L)~\footnote{ROUGE evaluates semantic coverage of key permission verification components, including resources, action mappings, and conditional statements.}~\cite{ROUGE} and 
ChrF++~\footnote{ChrF++ measures character-level similarity for code structure accuracy}~\cite{popovic2017chrf++}.}

\blue{As summarized in Figure~\ref{fig:finetuning}, the fine-tuned model achieved substantial performance improvements across all metrics. The BLEU score increased by approximately 220\% (24.2 to 77.5), ROUGE-2 by over 100\% (38.8 to 77.8), and ChrF++ by 124\% (37.4 to 83.9), with consistent improvements observed in other metrics.} These results demonstrate that the proposed approach can reliably generate high-quality verification logic that aligns with the complex permission requirements of serverless platforms and validate the effectiveness of fine-tuning in this context.

\begin{figure*}[t!]
    \centering
    \includegraphics[width=0.98\textwidth]{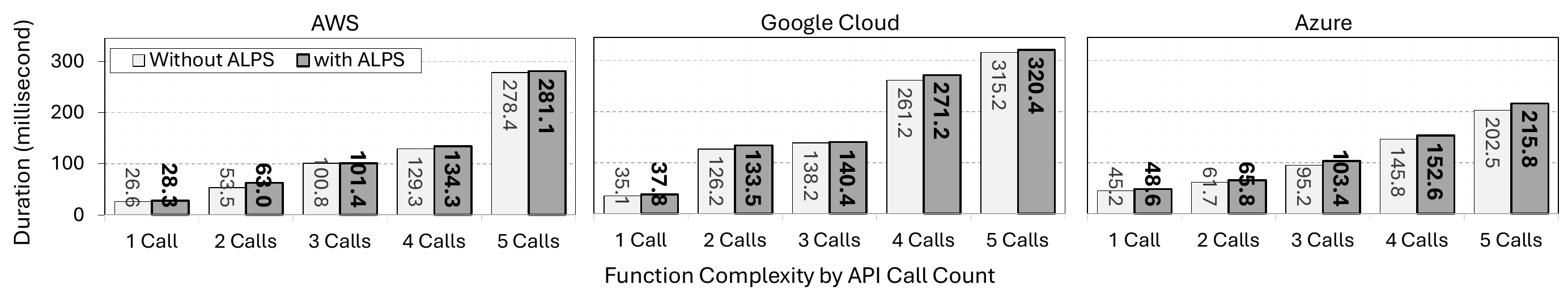}
    \caption{The results of the duration with and without \ourtool{} under different numbers of service calls in serverless functions across AWS, Google Cloud, and Azure.}
    \label{fig:overhead}
\vspace{-1.4em}
\end{figure*} 

\begin{table}[t]
\centering
\scriptsize
\renewcommand{\arraystretch}{1.2}
\caption{Cost comparison by cloud provider before and after applying \ourtool{} (\(\times 10^{7}\) USD) }
\label{table:cost}
\resizebox{\linewidth}{!}{%
\begin{tabular}{p{0.9cm}|l|c|c|c|c|c}
\Xhline{2.8\arrayrulewidth} & \textbf{\ourtool{}} & \textbf{1 Call} & \textbf{2 Calls} & \textbf{3 Calls} & \textbf{4 Calls} & \textbf{5 Calls} \\
\hline \hline
\multirow{2}{*}{AWS} 
  & without & 0.56 & 1.12 & 2.10 & 2.70 & 5.80 \\
  & with    & 0.59 & 1.31 & 2.11 & 2.80 & 5.86 \\
\hline
\multirow{2}{*}{\makecell{Google\\Cloud}} 
  & without & 7.69 & 27.64 & 30.26 & 57.21 & 69.07 \\
  & with    & 8.28 & 29.23 & 30.74 & 59.39 & 70.20 \\
\hline
\multirow{2}{*}{Azure} 
  & without & 0.90 & 1.23 & 1.90 & 2.92 & 4.05 \\
  & with    & 0.97 & 1.32 & 2.07 & 3.05 & 4.32 \\
\Xhline{2.8\arrayrulewidth}
\end{tabular}}
\vspace{-2em}
\end{table}

\textbf{Performance Overheads.}
This experiment quantitatively evaluates the impact of verification logic, inserted through code reconstruction by our system, on the execution time and cost of serverless functions. \blue{The functions were categorized by the number of service calls per function across AWS Lambda~\cite{awslambda}, Google Cloud Functions~\cite{gcpfunction}, and Azure Functions~\cite{azurefunction}.}
Each function was executed 100 times in a warm-start environment with identical regions and resource configurations. As shown in Figure~\ref{fig:overhead}, AWS Lambda exhibited the lowest overhead (1.7 ms for one service call, 2.7 ms for five), followed by Google Cloud (2.7 ms and 5.2 ms) and Azure (3.4 ms and 13.3 ms). Average overheads were 3.9 ms, 5.5 ms, and 7.2 ms, respectively, all remaining below 10\% across platforms.

Based on the measured execution times, we calculated the cost increase rates for each cloud platform using their respective pricing models. As shown in Table~\ref{table:cost}, execution costs (scaled to $\times 10^{7}$ USD) increased on average by 5.5\% for AWS, 4.1\% for Google Cloud, and 7.0\% for Azure. AWS showed the greatest variation, ranging from 0.5\% to 16.9\%, due to the interaction between its memory-based pricing and function complexity. Google Cloud showed relatively consistent cost increases driven by its cumulative vCPU and memory pricing. Azure exhibited steady cost increases proportional to overhead under its execution time pricing model.

Consequently, the performance and cost overhead introduced by our system remain within single-digit percentages, which is acceptable given the robust least-privilege security it provides. These results confirm that the approach offers a practical balance between security and efficiency in real-world serverless environments.

\section{Related Work}
\label{s:related}

\textbf{Serverless Permission Management and Flow Control.}
 
\blue{Recent research on serverless environments has focused on managing permissions and controlling information flow to enhance security, though most depend on specific languages or platforms~\cite{gupta2025growlithe, d2024automatically, polinsky2024grasp, alpernas2018secure, datta2020valve}. 
Gupta et al.~\cite{gupta2025growlithe} proposed a policy-based compliance verification tool for AWS and Google Cloud that validates whether serverless applications adhere to developer-specified policies. However, it fundamentally assumes developers have already manually configured all permissions according to the principle of least privilege, requiring them to examine the Application Dataflow Graph (ADG) and write policies using complex Datalog syntax for each edge, thereby placing substantial manual burden on developers.
In log-based approaches, D'Souza et al.~\cite{d2024automatically} proposed a tool that automatically reduces AWS IAM policies through access log analysis. However, it requires data collection periods and is dependent on logs. Polinsky et al.~\cite{polinsky2024grasp} identified potential attack paths through graph-based policy analysis for serverless, yet they did not provide any countermeasures. Alpernas et al.~\cite{alpernas2018secure} proposed dynamic information flow control for serverless, but it is limited to JavaScript and requires modifying function code. Datta et al.~\cite{datta2020valve} proposed information flow control for function workflows through network-level taint tracking, but it is specialized for the OpenFaaS only, which limits its generality.}

\blue{Unlike previous works, \ourtool{} offers a vendor-agnostic framework that automatically extracts least privilege permissions through serverless static analysis, supporting multiple programming languages and cloud platforms. It performs three levels of granular permission extraction at the service, object, and entity levels across AWS, Google Cloud, and Azure, automating permission management by generating deployable least privilege policies directly from function code.}

\textbf{Runtime Monitoring and Dynamic Security Enforcement.} 
In serverless environments, runtime security research mainly focuses on post-incident detection and auditing. ~\cite{shin2025bambda, shen2022gringotts, alpernas2021cloud, datta2022alastor, satapathy2023disprotrack} 
Shin et al.~\cite{shin2025bambda} proposed a dynamic security framework that detects and blocks privilege abuse and chain function call attacks in real time in serverless environments, but users must specify resources in advance for verification. Shen et al.~\cite{shen2022gringotts} developed a performance monitoring system that detects internal Denial-of-Wallet attacks in serverless computing. However, it only provides attack detection and does not actually block attacks, which allows them to continue after detection. Alpernas et al.~\cite{alpernas2021cloud} proposed a monitoring and debugging system that detects runtime property violations in serverless applications. However, it only records violations in logs after the fact without performing real-time blocking. Additionally, Datta et al.~\cite{datta2022alastor} proposed provenance-based auditing for attack path reconstruction, and Satapathy et al.~\cite{satapathy2023disprotrack} proposed distributed provenance tracking; however, both focus solely on post-attack analysis. 
Meanwhile, LLM-based approaches for security automation are still in the early stages in serverless environments~\cite{wen2025llm, arun2025llms}. 
\blue{These are limited to Wen et al.~\cite{wen2025llm}'s work on detecting errors in AWS configuration files and Arun et al.~\cite{arun2025llms}'s exploratory research on component generation.}

While existing approaches are reactive, detecting security violations only after they occur, \ourtool{} proactively prevents such violations in real time, blocks unauthorized service calls, and continuously enforces the principle of least privilege by automatically reevaluating code and policy changes using fine-tuned LLM-based adaptive code integration.

\section{Conclusion}
\label{s:conc}
This paper presents \ourtool{}, a framework that integrates large language models with AST-based static analysis to automatically extract least privileges and insert dynamic verification logic in serverless environments. Unlike prior graph- or log-based methods, it provides continuous enforcement of the principle of least privilege while maintaining minimal runtime and cost overhead. Experiments demonstrate high accuracy in privilege extraction, effective verification logic generation, and strong compatibility across major cloud providers and programming languages. 
\blue{By offering a practical, lightweight solution for security automation, this work provides a foundation for advancing serverless security.
The authors have made their code publicly accessible at \cite{alps} and their data at \cite{alpsdataset}.}

\section*{Acknowledgment}
\blue{This work was supported by the National Research Foundation of Korea (NRF) funded by the Korean Government (MSIT) under Grant RS-2025-16069415.}


\balance
\bibliographystyle{IEEEtran}
\bibliography{reference}

\end{document}